# Measurement of the absorption spectrum of $H_2O^+$ in the visible region

ZHUANG Hua, CHEN Yangqin, JI Wenhai, WU Shenghai, BI Zhiyi, LIU Yuyan & MA Longsheng

Key Laboratory of Optical and Magnetic Resonance Spectroscopy, Department of Physics, East China Normal University, Shanghai 200062, China
Correspondence should be addressed to Chen Yangqin (e-mail: yqchen@publicl.sta.net.cn)

**Abstract** Together with the 74 lines belonging to (0,9,0)-(0,0,0) band, the high-resolution absorption spectrum of $H_2O^+$ $\tilde{A}^2A_1 - \tilde{X}^2B_1$ system was observed in the visible region of $16680-17300$ cm$^{-1}$ using optical heterodyne magnetic rotation enhanced velocity modulation spectroscopy for the first time, which verifies the high sensitivity and high signal to noise ratio (S/N) of this technique.

**Keywords: $H_2O^+$, optical heterodyne technique, magnetic rotation spectroscopy, velocity modulation spectroscopy.**

$H_2O^+$ plays an important role in the study of the astrophysics, chemical physics and biophysics, as well as the study of the origin of life and searching living species on the other stars. Furthermore, the observation of $H_2O^+$ would be the only way of detecting the existence of the neutral $H_2O$ in the visible region, because $H_2O$ has no emission spectrum in the visible region due to the predissociation of its electronic excited states. In 1976, Lew[1] reported the observation of the emission spectrum of $H_2O^+$ in the visible region using a conventional grating spectrograph and performed a detailed analysis including the rotational transitions. However, under the same condition, he failed in getting the absorption spectrum. Due to much lower densities of molecular ions generated in a conventional discharge, and the interference of very intense lines from the neutral species occurring in the same region, especially, due to the very short lifetime and active chemical characteristics of molecular ions, it makes absorption spectroscopy more difficult to achieve.

In 1983, Gudman et al.[2] developed the velocity modulation spectroscopy (VMS) technique, which has been proved to be an effective method for selective detecting the spectra of molecular ions. Later, Das[3] and Huet et al.[4] obtained the high-resolution absorption spectrum of $H_2O^+$ in the near-infrared and infrared region using VMS. However, a serious problem in VMS is a coherent background noise arising from the glow emission and/or the electric pickup of the discharge to the detection circuitry. When the measurement of VMS is taken in the visible region, the amplitude fluctuation of dye laser at the velocity modulation frequency can also be a dominant noise to limit the measuring sensitivity. Despite of all the efforts such as using the dual beam subtraction technique[5] and double-modulation technique[6], it is still difficult to eliminate both the interference and the fluctuation noise completely. Recently, we developed an improved VMS technique named optical heterodyne magnetic rotation enhanced velocity modulation spectroscopy[7,8] (OH-MR-VMS) and applied it to detecting the absorption spectrum of the $\tilde{A}^2A_1 - \tilde{X}^2B_1$ system of $H_2O^+$ in the visible region for the first time. In this note, we reported the experimental results and preliminary analysis of 74 transitions which belong to the $\Sigma$ and $\Delta$ subband of (0,9,0)-(0,0,0) band.

## 1 Experiment

To improve the sensitivity of VMS in the measurement of the visible region, we combined the optical heterodyne detection technique and magnetic rotation spectroscopy (MRS) with the VMS. Optical heterodyne detection was implemented by phase modulation of the incident laser beam with an electro-optical modulator (EOM) driven at 480 MHz. The modulated laser beam passed through an absorption sample cell and then fell on an avalanche photodiode detector (APD) to obtain the rf beat note at 480 MHz which was generated by the laser carrier and its two sidebands. The laser fluctuation noise could be reduced to as low as its limit given by shot noise at such a high detected frequency. The modulation frequency in the experiment was chosen as 480 MHz to match the half-Doppler linewidth, and thus the maximum beat signal could be obtained without broadening linewidth.

$H_2O^+$ is a paramagnetic molecule with strong Faraday effect, which will alter polarization of the incident light. The absorption cell was surrounded by a solenoid, which provided a dc longitudinal magnetic field. A pair of nearly crossed polarizers with extinction ratio $5\times 10^{-5}$ placed on both sides of the cell was used to obtain the signal enhanced by the magnetic rotation effect. The combination of magnetic rotation spectroscopic technique not only greatly improves the S/N, but also presents the advantage for selective detection of paramagnetic species.

The detail description of the experimental setup for OH-MR-VMS has been given in ref. [7]. In the present experiment, a ring dye laser (Coherent 899-29), pumped by a compact cw diode-pumped Nd:YVO$_4$ laser system (Verdi 10) was used to provide a tunable radiation with single-mode output power of higher than 300 mW. The autoscan function in a wide range and the precise determination of laser wavelength in real time by an attached wavemeter are provided by the laser system. An I$_2$ reference absorption cell was used for calibration of the absolute wavenumber, the accuracy was better than 0.006 cm$^{-1}$, and the relative precision better than 0.001 cm$^{-1}$.

The discharge cell was made of a 40-cm-long quartz pipe with an internal diameter of 1 cm. The mixture of $H_2O$ vapor (25 Pa) and He gas (665 Pa) flowed continuously through the cell with the discharge current 250 mA peak-to-peak at 37 kHz. The transmitting light through the cell was focused onto the APD detector attached a narrow band-frequency tuned preamplifier at 480 MHz, which further reduced the discharge interference noise at 37 kHz. The output from the detector was demodulated by a double balance mixer (DBM) at the frequency of 480 MHz and then by a lock-in amplifier at the velocity modulation frequency of 37 kHz. Finally, data were acquired and processed by a computer. The experimental apparatus is such a kind of system with double-modulation and double-demodulation. The sensibility of this absorption spectroscopic technique has been greatly improved due to the combination of those three kinds of zero-background spectroscopic techniques.

## 2 Result and discussion

Using OH-MR-VMS technique, we obtained 74 lines in the region of 16700—17300 $cm^{-1}$. The line positions are the same as Lew's results[1]. However, in comparison with somewhat low resolution of the emission spectrum technique, in the experiment, the frequency stabilized laser source with the single mode was used and the wavelength calibration was simultaneously recorded according to the $I_2$ spectra while the scanning laser, the accuracy of the wavenumber were determined as better than 0.006 $cm^{-1}$. All these 74 lines belong to the $\Sigma$ and $\Delta$ subbands of (0,9,0)-(0,0,0) vibational band of $H_2O^+$ $\tilde{A}\,^2A_1 \sim \tilde{X}\,^2B_1$ electronic transition. In the emission spectra of Lew[1], some strong lines, which were not assigned, mostly came from perturbation of the neutral species. However, due to the profit of VMS, all spectra of those interfering species disappeared, except those of $H_2O^+$ that could be detected in our measurement.

Fig. 1 shows the portion of the absorption spectra of the $\Sigma$ subband of $^PQ_{1,N}$ band. The line shape presents the second derivative in an approximation due to common contributions from the optical heterodyne detection and velocity modulation technique. In contrast with the observation in the near infrared region by Heut et al.[4], the absorption cell which they used was $2\times 1$ m in length and the integral time constant was 1 s, while that in our experiment was only 0.4 m in length and 0.3 s, respectivively. Moreover, the absorption was weaker in the visible region than that in the infrared due to the Frank-Condon principle, the signal-to-noise ratio in our experiment was still a little better than those by Heut et al.

As shown in fig. 1, due to the statistical weights exchanging two identical hydrogen nuclons, the intensity of the lines changes alternatively in the ratio 3∶1 with the rotational quantum number $N$. According to the interaction of the electronic spin and molecular rotation, with a fixed $N$ value the line splits into two lines to form a pair of spin components F1($J = N + 1/2$) and F2($J = N - 1/2$) with the opposite phase due to the magnetic rotation effect, therefore, this is very helpful to confirm the assignment of the spectral lines.

Figs. 2 and 3 show the variation of the signal intensity with increase of the magnetic field strength in the measurement of the line near 17237.9 cm$^{-1}$. The intensity grows linearly with the magnet field within the magnetic intensity range of 40 mT.

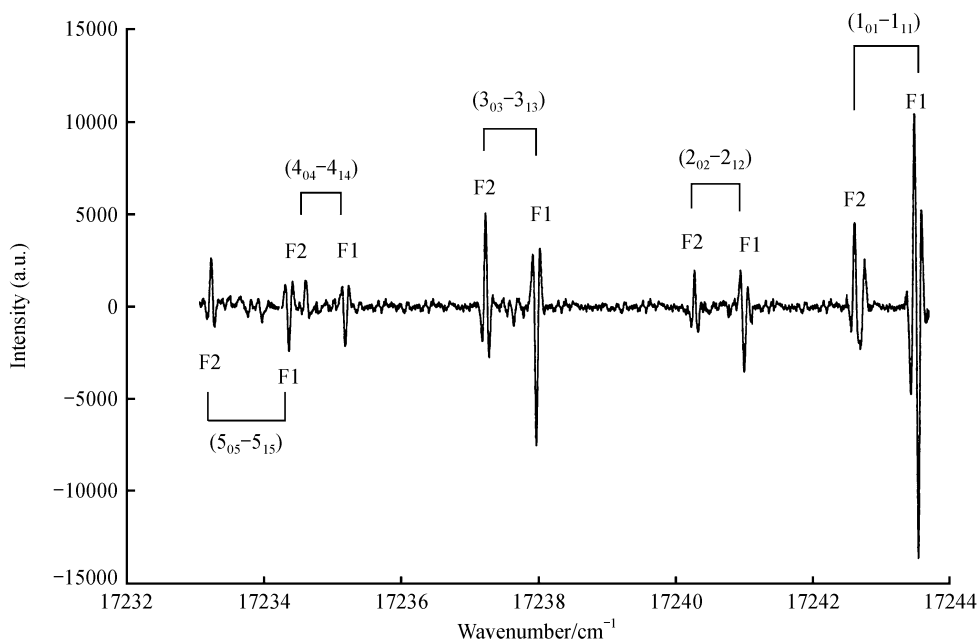

Fig. 1. Portion of the the spectra of $^PQ_{1,N}$ band in the $\tilde{A}\,^2A_1 - \tilde{X}\,^2B_1$ of $H_2O^+$.

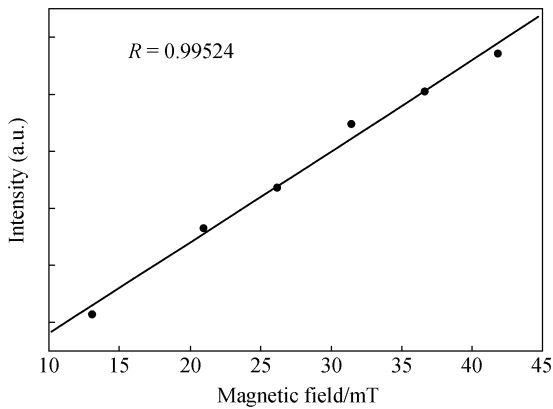

Fig. 2.  The spectral intensity varies with magnetic field.

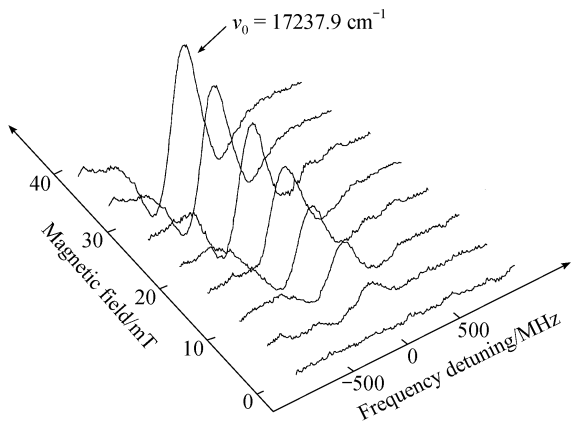

Fig. 3.  The signal in the different magnetic field.

In our measuement, by introducing the optical heterodyne and MRS technique, the fluctuation noise from the laser source and the coherent discharge noise could be ignored, while the shot noise was considered as a dominant one. In this situation, the minimum detectable sensitivity for $H_2O^+$ can be estimated as follows if $S/N = 1$[9]:

$$\Delta\delta_{min} = \frac{4}{f(B) \cdot J_0(M) \cdot J_1(M)} \cdot \sqrt{\frac{e \cdot \Delta f \cdot F}{k \cdot P_0}},$$

where the $f(B)=1+k_B \cdot B$ (here $k_B = 0.93$/mT) is the enhancement factor of magnetic field which has been concluded from the experimental results (see fig. 2.); $J_i(M)$ are $i$th Bessel components for the carrier and sidebands of the modulated laser beam with the modulation degree of $M$, the maximum value of $J_0(M)J_1(M)$ is equal to 0.34 when $M = 1.1$, $e = 1.9 \times 10^{-19} C$ is the charge of electron, $\Delta f = 1/2\pi T$ is the detection bandwidth where $T$ is the lock-in integral time constant, $F = G^{\chi} = 3.2$ is the excess noise factor of APD detector where $G = 10$ is the gain of the detector and $\chi =$ 0.5 is the excess noise figure. $k = \dfrac{e\eta}{\hbar\omega}$ denotes the photo current per unit light power, $\eta = 0.77$ is the quantum efficiency. So the achievable relative minimum absorptivity in our system is about $2.9 \times 10^{-10}$. Actually, the practical sensitivity is always lower than this estimation because of the existence of residual amplitude modulation (RAM) from the imperfect frequency modulation of the EOM. However, the result agrees well with the exeriment.

The analysis of $H_2O^+$ spectra is still under processing. $H_2O^+$ is an asymmetry top molecule, and its ground state $\widetilde{X}$ and the first excited electronic state $\widetilde{A}$ are two Renner-Teller components of a degenerated $^2\Pi_u$ state in a linear configuration. It reflects the interaction between the electronic motion and the nuclear vibration of the molecule. Under this situation, the Born-Oppenheimer approximation breaks down. In 1993, an *ab initio* calculation was performed by Brommer et al.[9], who considered the Renner-Teller effect to deduce the three dimension potential function of $^2A_1$-$^2B_1$ system of $H_2O^+$, $D_2O^+$ and $HDO^+$. The result of theoretical calculation agreed well with that of the experiment. To further verify those kinds of theoretical problems, the precise measurement of $H_2O^+$ spectra with high sensitivity and high resolution will be of great significance.

**Acknowledgements**  The authors would like to thank Prof. Ding L. E. and Yu Shanshan for the use of the laser. This work was supported by the National Natural Science Foundation of China (Grant No. 19974011) and the Laboratory of Magnetic Resonance and Atomic and Molecular Physics (Grant No. 991516).

### References

1. Lew, H., Electronic spectrum of $H_2O^+$, Can. J. Phys, 1976, 54: 2028.
2. Gudeman, C. S., Saykally, R. J., Velocity modulation infrared laser spectrocopy of molecular ions, Ann. Rev. Phys. Chem., 1984, 35: 387.
3. Das, B., Farley, J. W., Observation of the visible absorption spectrum of $H_2O^+$, J. Chem. Phys., 1991, 95: 8809.
4. Huet, T. R., Bachir, I. H., Destombes. J. L. et al., The transition of $H_2O^+$ in the near infrared region, J. Chem. Phys., 1997, 107: 5645.
5. Nesbitt, D. J., Petek, H., Gudeman, C. S. et al., A study of the $v_1$ fundamental and bend-excited hot band of $DNN^+$ by velocity modulation absorption spectroscopy with an infrared difference frequency laser, J. Chem. Phys., 1984, 81: 5281.
6. Lan, G., Tholl, H. D., Farley, J. W., Double-modulation spectroscopy of molecular ions: Eliminating the background in velocity-modulation spectroscopy, Rev. Sci. Instrum., 1991, 62: 944.
7. Wang, R., Chen, Y., Cai, P. et al., Optical heterodyne velocity modulation spectroscopy enhanced by a magnetic rotation effect, Chem. Phys. Lett., 1999, 307: 339.
8. Wang, R., Chen, Y, Cai, P. et al., A new spectroscopy technique of detecting molecular ions, Optics Transaction(in Chinese), 2000, 20: 14.
9. Luo, M., Bi, Z., Cai, P. et al., Optical heterodyne Stark modulation spectroscopy, Chem. Phys. Lett., 2000, 327: 171.
10. Brommer, M., Weis, B., Follmeg, B. et al., Theoretical spin-rovibronic $^2A_1(\Pi u)$–$^2B_1$ spectrum of the $H_2O^+$, $HDO^+$, and $D_2O^+$ cations, J. Chem. Phys., 1993, 98: 5222.